Title: ATP consumption of eukaryotic flagella measured at a single-cell level

Authors: Daniel T.N. Chen[1], Michael Heymann[1,2], Seth Fraden[1], Daniela Nicastro[3,*], and Zvonimir Dogic[1,*]

[1] Martin Fisher School of Physics, Brandeis University, 415 South St., Waltham, MA 02454, USA
[2] Graduate Program in Biophysics and Structural Biology, Brandeis University, 415 South Street, Waltham, MA 02454, USA
[3] Department of Biology, Brandeis University, 415 South St., Waltham, MA 02454, USA

*Address correspondence to nicastro@brandeis.edu, zdogic@brandeis.edu

**Abstract:**

The motility of cilia and flagella is driven by thousands of dynein motors that hydrolyze adenosine triphosphate (ATP). Despite decades of genetic, biochemical, structural and biophysical studies, some aspects of ciliary motility remain elusive, such as the regulation of beating patterns and the energetic efficiency of these nanomachines. Here, we introduce an experimental method to measure ATP consumption of actively beating axonemes on a single-cell level. We encapsulated individual sea urchin sperm with demembranated flagellum inside water-in-oil emulsion droplets and measured the axoneme's ATP consumption by monitoring fluorescence intensity of a fluorophore-coupled reporter system for ATP turnover in the droplet. Concomitant phase contrast imaging allowed us to extract a linear dependence between the ATP consumption rate and the flagellar beating frequency, with ~$2.3 \times 10^5$ ATP molecules consumed per beat of a demembranated flagellum. Increasing the viscosity of the aqueous medium led to modified beating waveforms of the axonemes and to higher energy consumption per beat cycle. Our single-cell experimental platform provides both new insights into the beating mechanism of flagella and a powerful tool for future studies.

# Introduction:

Cilia and flagella are ubiquitous organelles for motility and sensing in eukaryotic organisms. At the core of each flagellum is an axoneme, a highly conserved superstructure comprised of 9 doublet microtubules surrounding a central pair of singlet microtubules in a "9+2" pattern. Attached to the doublets are thousands of dynein motors that - fueled by ATP hydrolysis - walk on microtubules, inducing sliding between neighboring doublets, which in turn is converted into flagellar bending by constraints that limit interdoublet sliding. Over the past 50 years, detailed genetic (1, 2) and biochemical studies (3) in conjunction with advances in ultrastuctural imaging (4, 5) have uncovered the key structural components of the axoneme. However, due to the structural complexity of the axoneme a consensus mechanism for how all these components work together to generate high frequency beating patterns remains elusive (6, 7). Many candidate mechanisms for regulation of local dynein activity have been proposed, including regulation by curvature (8), normal force (9), and sliding control (10, 11). In order to test predictions derived from these models, precise experimental data, ideally at a single-cell level, are needed.

Previous studies of demembranated flagella by Brokaw (12-14) and Gibbons (15) found that axonemal ATP consumption occurs in both active (beating) and non-active flagella, albeit at different consumption rates. These measurements were performed on a population-level in "bulk" by "pH-stat" methods, that is by detecting pH changes that were caused by ATP hydrolysis in a solution containing ~$10^9$ cells with demembranated flagella. However, such solutions are typically polydisperse mixtures of cells with active and non-active flagella, because under demembranation and reactivation conditions usually less than 100% of the cells are fully motile and the percentage of immotile and fragmented cells increases with viscosities higher than that of water (13). Therefore, previous measurements were likely averages over different levels of energy consumption, resulting in a mean ATP consumption rate that is lower than the rate of fully active flagella, posing major limitations to the interpretation of bulk ATP consumption rates and the applicability of bulk methods.

To avoid bulk averaging and allow direct correlation between an axoneme's beating and energy consumption, we optimized sample preparation protocols and develop an experimental platform that simultaneously permits direct measurement of dynein-dependent ATP consumption and waveform kinematics from a single cell encapsulated in a water-in-oil emulsion droplet (Fig. 1). Most importantly, this direct correlation between the energetic measurement and the motility of an individual flagellum circumvents the challenge of achieving 100% reactivation and permits a wider experimental range, such as measurements in high viscosity solutions, on immotile fragments, and at different ATP concentrations, including during the transition from beating to non-beating states. In addition to benefiting future studies, this single-cell method has already provided novel information about ATP consumption rates, beating frequencies, waveforms, and their dependence on control parameters such as fluid viscosity, which may prove to be pivotal for discrimination between proposed regulatory mechanisms of flagellar motility.

**Materials and Methods:**

*Sperm Harvesting:* Lytechinus pictus sea urchins were purchased from Marinus Scientific (Long Beach, CA) during their fertile months between April–October and were kept in a sea water aquarium at 15.6 °C for typically ~1-14 days prior to use. Spawning of male sea urchins was induced by injection of ~1 mL of 0.5 M KCl into the perivisceral cavity. Undiluted sperm-solution was collected with a pipette and stored on ice until used in the experiment. Typically, samples were used within ~8 hours of spawning.

*Sperm Demembranation and Reactivation:* Sperm were demembranated using Triton-X according to a previously published protocol (15). Briefly, just prior to use 10 μL of harvested sperm solution was gently diluted into 50 μL of calcium-free artificial sea water (ASW) and left on ice for 2 min (16). Next 1 mL of extraction solution (ES) containing 200 mM K-acetate, 2 mM $MgCl_2$, 0.5 mM EGTA, 0.1 mM EDTA, 1 mM DTT, 10 mM Tris-HCl (pH 8.3), and 0.04% w/v Triton-X was added to the sperm-ASW solution, gently mixed, and left on ice for 2 min. Finally, 2 μL of the sperm-ASW-ES solution was pipetted into reactivation solution (RS) containing 200 mM K-acetate, 2 mM $MgCl_2$, 0.5 mM EGTA, 0.1 mM EDTA, 1 mM DTT, 20 mM Tris-HCl (pH 8.3).

In addition to the RS we added components of an ATP regeneration system that was coupled to a Nicotinamide adenine dinucleotide (NADH)-based optical readout system for ATP-concentration (17). The regeneration system rephosphorylates the ADP that is generated when dynein motors hydrolyze ATP, thus maintaining a stable concentration of ATP. The pyruvate byproduct is coupled downstream to a redox reaction which turns NADH, which is fluorescent at a wavelength of 340 nm, into $NAD^+$ which is non-fluorescent at 340 nm (Fig. 1C). The ATP regeneration system contained ATP at the desired concentration (between 1 to 200 μM), 10 μL 60 mM phosphoenolpyruvate (PEP), 3.5 μL pyruvate kinase/ lactic dehydrogenase (PK/LDH) (Sigma-Aldrich, St. Louis, MO), and 7.0 μL of 6 mM NADH (Sigma-Aldrich). ATP, DTT, NADH, and PEP were freshly thawed from frozen aliquots and added to the RS just prior to experiments. The final volume of sperm-RS-ATP solution was 200 μL. For the experiments where viscosity was varied, we added 0.5 – 2.0% w/v of 4000 cP methyl cellulose powder (Sigma-Aldrich) to 1 mL of RS followed by gentle agitation overnight at 4°C.

*Chlamydomonas Strains and Culture Conditions:* We used two *Chlamydomonas reinhardtii* strains: Wild-type (WT) strain (CC-125, 137c mt+, obtained from *Chlamydomonas* resource center, University of Minnesota, St. Paul, MN) and a normal-length, paralyzed mutant WS4 strain (pf28, pf30, ssh1, obtained from W. Sale lab, Emory University) missing all outer dynein arms and the inner dynein arm subform I1 (18, 19). In addition to the missing dyneins, WS4 also has a third mutation ssh1, an allele that encodes a flagella-associated protein FAP234 believed to be part of a polyglutamylation complex involved in regulation of ciliary motility (20). Cells were grown in liquid Tris acetate/phosphate medium with a light/dark cycle of 16:8 h (21).

*Chlamydomonas Axoneme Preparation:* Deflagellation of *C. reinhardtii* cells was induced by pH shock (22). Flagella were collected by subsequent centrifugation. To prepare axonemes, flagella were demembranated with 0.1% Igepal CA-630 (Sigma-Aldrich, St. Louis, MO) in HMDEK (30 mM Hepes, 5 mM $MgSO_4$, 1 mM DTT, 1 mM EGTA, and 50 mM potassium acetate, pH 7.4) and centrifuged to remove the membrane and matrix fraction. Purified axonemes were resuspended in 30 mM Hepes, pH 7.4, 25 mM KCl, 5 mM $MgSO_4$, 1 mM EGTA, and 0.1 mM EDTA and either used within 24 h or diluted in glycerol to 70%, plunge frozen in liquid nitrogen, and stored at -80 °C.

*Chlamydomonas Axoneme Reactivation:* Following (23), axonemes were thawed just prior to use and resuspended to a density of ~$10^8$ /mL in HMDEKP (30 mM Hepes, 5 mM $MgSO_4$, 1 mM DTT, 1 mM EGTA, and 50 mM potassium acetate, 1% PEG, pH 8.3) along with the components of the ATP regeneration system identical to sperm axonemes described previously. All *Chlamydomonas* experiments were carried out at [ATP] = 50 μM. The protocol for droplet encapsulation of *Chlamydomonas* axonemes is identical to that of sperm axonemes as described below.

*Droplet Encapsulation:* Water-in-oil emulsions, in which the aqueous phase was the sperm-RS-ATP solution, were formed by gentle agitation of 0.5 μL of sperm-RS-ATP solution added to 2.0 μL of perfluorinated oil HFE-7500 (3M, St. Paul, MN) in a microcentrifuge tube. To stabilize the droplets, the oil contained 2% v/v of a PFPE-PEG-PFPE surfactant (RAN Biotechnologies, Beverly, MA), which has been shown to be highly compatible with proteins (24, 25). After the emulsification, the droplets were imbibed by capillary pressure into glass capillary tubes with rectangular cross-section of dimensions 20 μm height by 200 μm width and 5 cm length (VitroCom, NJ # 5002S-050). When loaded into the capillaries, emulsion droplets assumed a flattened disk profile with the radius ranging from 50–120 μm (Fig. 1A), corresponding to an encapsulation volume of 0.1-1 nL. To prevent evaporation and flow, the ends of the capillary were sealed with acrylic nail polish (Sally Hansen; Coty Inc., New York, NY). Droplets typically contained between 0 to 5 sperm cells, with varying degrees of beating activity. For the experiments in this paper, we focused exclusively on droplets containing a single sperm cell with either active or inactive flagellum (Fig. 1B), or empty droplets as controls, respectively.

Due to strong hydrodynamic attraction to surfaces, most sperm cells swam near the droplet's upper and lower surfaces in circular trajectories with radii comparable to the length of the sperm (Movie S1). This circular swimming behavior does not seem to be induced by the presence of the droplet, because unencapsulated sperm in microscope chambers exhibited circular trajectories with similar radii. In addition, we did not observe any systematic dependence of ATP consumption rate or beating frequency on the droplet size (Fig. S2).

*Microscopy Protocol:* Emulsion droplets were imaged using an automated inverted light microscope (Nikon Ti; Nikon, Tokyo, Japan) using a phase contrast objective (40X, 0.75 NA). Phase contrast was used to assess the number, quality and activity state of the cells, i.e. if the flagellum was beating. We considered only those droplets containing a single sperm cell with fully propagating, regular waveforms, while discarding those with multiple or damaged sperm with irregular and partially beating waveforms. Once a suitable droplet was identified, flagellar beating waveforms were recorded at 10-100 frames per second with an Andor Neo CMOS or Phantom v9.0 camera (Movie S1). Subsequently, the droplet fluorescence was imaged using a filter set consisting of 340 ± 30 nm notch excitation filter, 380 nm short pass dichroic, and 400 nm long pass emission filter. Due to its high UV transmission we used FL Plan 40X objective. Illumination from a mercury halide light source (Lumen X-cite 120; Excelitas Technologies, Waltham, MA) was used to excite the NADH in the droplet and fluorescence was recorded with 16-bit pixel depth, 4x binning, and 2 frames per minute (Andor Clara; Andor Technology Ltd., Belfast, UK). To minimize photobleaching, the excitation source was shuttered. Throughout the course of the imaging, typically lasting ~30 minutes, the sperm's motility was periodically confirmed using phase contrast.

Dark-field images of sperm swimming near coverslip (Fig. 3D,E) were obtained using a Nikon dark-field condenser under low-pass filtered ($\lambda$ < 640 nm) illumination from a Mercury Halide light source (Lumen X-cite 120) and imaged using an oil-immersion objective (100X, 1.3 NA) coupled to a demagnifying lens (0.55X). The images were recorded using a high-speed CMOS camera (Phantom v.9.0; Vision Research Inc., Wayne, NJ ) at 100 frames per second.

*Image Analysis:* Droplet fluorescence images were analyzed using custom routines written in IDL (Exelis Inc., Mclean, VA). Intensity thresholding was used to determine the droplet size. Images of fluorescent droplets had a nearly "flat-top" intensity profile due to their disk-like compression in the rectangular capillaries. Circles were fit to the binary image and the center-of-mass coordinates, as well as the radius of the droplet were extracted. Subsequently, the ratio of the mean intensity in the circular droplet region was determined and divided by the mean intensity of the background region. These ratiometric intensities (Fig. 2C), unlike absolute intensities, are robust against multiplicative intensity fluctuations. To relate the ratiometric intensity to the actual ATP concentration in the droplet, we performed calibration measurements by adding known amounts of ADP instead of ATP to the coupled reactivation/ reporter system and obtaining calibration curves that showed a linear dependence of the ratiometric intensity on the concentration of ADP (Fig. S1). The linear dependence confirms the 1:1 stoichiometric dependence of [ATP] to [NADH]. Using the calibration curve, we first converted the raw ratiometric intensity to [ATP] in the droplet, then after multiplying [ATP] by the disk-like volume of the compressed emulsion, we obtained the absolute ATP consumption rate (in units of ATP molecules/second).

The sperm beating frequency was determined visually using frame-by-frame playback of the phase contrast movies to determine the time interval between repeating waveforms in the sperm's beat cycle. The reported beating frequency for each sperm is the average of five independent measurements. Measured in this way, variations in the beating arising from non-periodicity would be reflected in the variance of the measured beating frequency. A plot of the mean beating frequency over 5 independent measurements for each sperm is shown in Figure S3. From the size of the error bars relative to inter-sperm frequency deviations from each other, it is clear that the temporal resolution of our measurement is sufficient to resolve variations between different cells under identical experimental conditions.

## Results:

### Measurement of ATP consumption rates on a single-cell level

To measure ATP consumption rates we monitored the fluorescence signal of droplets, each containing a single sperm cell with beating flagellum. We found that the fluorescence of droplets encapsulating an actively beating sperm decayed significantly on a timescale of a few minutes (Fig. 2B). In comparison, a droplet with control solution containing boiled and thus enzymatically completely inactive sperms exhibited minimal signal decay on this timescale (Fig. 2A). This difference is illustrated in supplemental movie S2, which shows two equivalently sized droplets, of which one (left) was empty and the other (right) contained an actively swimming sperm. In latter droplet the measured fluorescence intensities decayed monotonically with time due to ATP turnover by the demembranated, beating flagellum. In experiments at different ATP concentrations we found a clear dependence of the rate of signal decay to the [ATP] (Fig. 2C). Brightfield phase contrast imaging showed also that at higher [ATP] the flagella were beating faster, which is likely the reason for a higher rate of ATP consumption. Sperm-free control droplets exhibited only a minimal decay of fluorescence signal due to photobleaching. To correct for this effect, the photobleaching contribution was subtracted from all measured droplet intensity decay rates.

Droplet fluorescence measurements on several different sperm were made for [ATP] spanning the range 0 to 100 μM. The raw ratiometric intensities were converted to absolute ATP consumption rates (in units of ATP molecules/second) using conversion factors derived from ADP calibration data (Fig. S1). Figure 3A shows a clear dependence of the ATP consumption rate on [ATP]. We find the dependence of the rate on [ATP] is well described by a Michaelis-Menten functional form, $V_{ATP} = V_{max}[ATP]/(K_M + [ATP])$, with $V_{max} = 6.2 \pm 1.7 \times 10^6$ ATP/s and $K_M = 51.7 \pm 24.8$ μM. Michaelis-Menten enzyme-substrate kinetics predicts saturation with a linear onset, as observed in our data, rather than sigmoidal onset indicative of positive cooperativity. This suggests that there is negligible cooperativity in ATP binding to dynein during flagellar beating. Previous bulk studies conducted on a different sea urchin species (*Colobocentrotus atratus*) and using "pH-stat" methods, whereby ATP consumption rates were determined as an average over millions of cells, have also obtained Michaelis-Menten-like kinetics for the ATP consumption rate vs [ATP], although with somewhat different values of $V_{max}$ and $K_M$ (15).

### ATP consumption rate and beat frequency show a linear relationship

Along with the ATP consumption rates, we concurrently measured the beating frequencies of the same sperm cells. The mean beating frequency as a function of [ATP] is shown in Figure 3B. Unlike the ATP consumption rate, which increases continuously from 0 following Michaelis-Menten kinetics, we did not observe sustained beating below a critical ATP concentration $[ATP]_0 = 5$ μM. At this [ATP] the critical beating frequency $V_0 = 0.5$ Hz. Below this concentration only sporadic, highly aperiodic beating occurs. The critical [ATP] required for the onset of sustained beating indicates that a critical fraction of the motors have to be simultaneously active in order to generate a macroscopic beating pattern. Above $[ATP]_0 = 5$ μM, the beating frequency increases with [ATP] following Michaelis-Menten-like scaling: $V_{ATP} = V_0 + V_{max}([ATP]-[ATP]_0)/(K_M + ([ATP] - [ATP]_0))$, with $V_{max} = 41.6 \pm 9.8$ Hz and $K_M = 141.6 \pm 35.4$ μM. The $[ATP]_0$ and subsequent Michaelis-Menten-like scaling for the beating frequency vs [ATP] are also in accord with previous bulk studies (15, 26).

To determine the number of ATP molecules consumed per beat, we plotted the ATP consumption rate against the beating frequency. The graph shows a linear relationship between ATP consumption rate and beating frequency (Fig. 3C). The slope of the resulting line yields an estimate for the number of ATP molecules consumed per beat: $(2.3 \pm 0.2) \times 10^5$ ATP/beat. Previous bulk measurements using sperm from the sea urchin *Strongylocentrotus purpuratus* also determined a linear dependence of ATP consumption rate on beating frequency, with a resulting slope of ~$1 \times 10^5$ ATP/beat, which is lower than our measurements (13).

## *In high viscosity medium the waveform and efficiency of axonemes change, but not the ATP consumption rate*

It has previously been shown that the beating waveform can vary significantly as a function of viscosity of the surrounding medium (26, 27). For this reason, we have examined the effects of increasing fluid viscosity on the ATP consumption rate and beating frequency. Solution viscosity was increased by addition of 0.5% w/v methyl cellulose, resulting in a viscosity ~8X that of the buffer, as determined by particle tracking microrheology (28). Even at this modest increase in viscosity, axonemes exhibited a strikingly different beating waveform. Compared to the buffer waveform, the higher viscosity waveform is more "compressed" to tight meanders, that is the flagella exhibit a sinusoidal shape with a smaller amplitude and shorter wavelength (Fig. 3D, E). Despite this significant difference in waveform, however, the flagella in higher viscosity fluid consumed ATP at the same rate as those placed in a pure buffer. High viscosity data are still well described by Michaelis-Menten kinetics, with $V_{max}$ and $K_m$ indistinguishable from the data obtained in low viscosity buffer (Fig. 3F).

While the ATP consumption rate of an axoneme was found to be independent of fluid viscosity, its beat frequency changed significantly. Overall the beat frequency is lower at higher viscosity, with increasing differences for higher [ATP]. Fitting the beating frequency dependence on [ATP] concentrations to Michaelis-Menten-like kinetics as before yields $V_{max}$ and $K_m$ that are roughly half of those measured for pure buffer (Fig. 3G). Combining measurements of ATP consumption rate and beating frequency at high viscosity, we plotted ATP consumption rate as a function of beat frequency and found a linear dependence as before, albeit with a higher slope corresponding to $(3.2 \pm 0.5) \times 10^5$ ATP/ beat (Fig. 3H), i.e. ~1.5x higher than in buffer. This increased energetic cost is perhaps not surprising, as the generation and maintenance of the highly curved waveforms at high viscosity likely require more elastic energy than the less curved waveforms at buffer viscosity.

*Inactive, non-beating flagella show varying residual ATP consumption rates*

In addition to actively beating axonemes, we also examined droplets that contained a single sperm with an inactive, non-beating flagellum. Figure S4 shows the ATP consumption rates for both active (same data as Fig. 3A) and inactive (non-beating) sperm as a function of [ATP]. Over the entire range of ATP concentrations, the inactive axonemes consumed ATP at ~30% of the active axoneme consumption rate, which is consistent with previous bulk studies utilizing mechanically broken *S. purpuratus* axonemes (13). Our measurements of droplets containing only cell fragments indicate that the sperm head without flagellum has negligible ATP consumption. Therefore, the measured ATP consumption stems from the non-beating axonemes, suggesting endogenous, movement-independent ATPase activity. Considering that there are thousands of dyneins with ATPase function, they are a likely source of ATP-hydrolysis, e.g. by dyneins that are thought to have a regulatory rather than a motility function (I1 dynein and some of the single-headed inner dyneins (29, 30)), or by uncoordinated, futile dynein activity. High frequency oscillations observed in non-beating, demembranated sea urchin axonemes have been attributed to such uncoordinated dynein activity (31).

An important advantage of the method developed here is that it enables measurements of ATP consumption of cells which transition from beating to non-beating flagella. Figure 4A shows the ATP consumption as a function of time for two different sperm cells with flagella that were beating at the beginning of the measurements, but then abruptly ceased movement during the course of the measurement. For both cells, the slope of the data decreases after the transition, indicating that the non-beating ATP consumption rate is lower than the beating ATP consumption rate. However, in contrast to the beating ATP consumption rate, the non-beating ATP consumption rate exhibited a larger difference between the two cells. This difference in residual ATP consumption rates could result from cell-to-cell variation in endogenous (movement-independent) ATP consumption and/or from variation in degree of mechanical damage that caused the beating to cease. Figure 4B shows all our data (N = 12) for the beating to non-beating transition as a function of [ATP]. We found that the ratio of non-beating to beating rate varied between 0.5 – 1.0, with a sharp peak in the distribution at ~0.85 (Fig. 4C). Interestingly, the 0.85 ratio found for freshly transitioned cells was systematically higher than the ratio of 0.3 found for inactive to active ATP consumption rates measured for cells that have never been observed to beat. This suggests that in freshly transitioned cells beating failure occurs via a mechanism that retains higher rates of residual movement-independent ATP consumption. We did not observe the ATP consumption rate of freshly transitioned cells to drop to 0.3 over the course of our measurements, which continued up to ~ 30 min after movement ceased. Instead, the inactive ATP consumption rate remained constant for both sperm that freshly transitioned and for sperm that never exhibited beating over the course of our measurements. Possible reasons for this variation in residual ATP consumption rates will be discussed below.

*Dynein is the major contributor to ATPase activity in both active and quiescent flagella*

Our single-cell measurements of the beating transition reveal that inactive axonemes can consume ATP at a rate which is nearly the same as the rate of active axonemes. This raises an important question of whether dynein is the major contributor to ATPase activity in the axoneme or whether other endogenous ATPases dominate ATPase activity, resulting in the lack of dependence on motility of the ATPase rate. In order to answer this question, we need to be able to distinguish between the ATPase activity of dynein from other ATPases in the axoneme. However, due to the lack of specific inhibitors for axonemal dynein and the unavailability of genetic modification options for sea urchin sperm, this line of investigation is a very challenging to pursue using the sea urchin sperm system. Instead, we elected to carry out single-cell measurements of ATPase activity using *Chlamydomonas reinhardtii* axonemes, which have the advantage of having a well-characterized library of axonemal protein mutants (1, 2). Specifically, we compared *Chlamydomonas* wild-type (WT), which has 12 outer dyneins (grouped in 4 outer dynein arms), and 8 inner dyneins (the dimeric I1 dynein and 6 single-headed inner dynein arms) per 96 nm repeat to a paralyzed mutant strain WS4, which constitutively lacks all 12 outer dyneins and 2 of the 8 inner dyneins, missing a total of 14/20 dyneins found in the WT axonemal repeat (18, 19).

We found that the ATP consumption rate for the WS4 axonemes is dramatically lower than both active and inactive WT axonemes (Fig. 5). The data decisively demonstrate that dynein motors are the major contributor to ATPase activity in the axoneme since, if ATPase activity were dominated by non-dynein axonemal ATPases, we would have observed no difference in ATPase rate between WT and dynein-deficient WS4 axonemes. Specifically, we found $(1.1 \pm 0.4) \times 10^5$ ATP/s for (paralyzed) WS4, $(4.7 \pm 0.7) \times 10^5$ ATP/s for the inactive WT, and $(9.7 \pm 0.4) \times 10^5$ ATP/s for actively beating WT axonemes. The ratio of WS4 to inactive WT ATPase rate is ~ 0.25, which corresponds well with the 0.3 expected from WS4 having only 6/20 dyneins of WT. In addition, our measured ATPase rates for active and inactive *Lytechinus* sperm, $(2.4 \pm 0.3) \times 10^6$ ATP/s and $(9.2 \pm 0.2) \times 10^5$ ATP/s, respectively, are about double of the corresponding active and inactive WT *Chlamydomonas* axonemes. This corresponds well to the expected ratio based on the fact that sea urchin sperm flagella are ~3.5 times longer than WT *Chlamydomonas* flagella, but contain only 16, rather than 20, dyneins per 96 nm repeat, because they have only two dyneins per outer dynein arm compared to three dynein heavy chains in *Chlamydomonas* (32); therefore, sea urchin axonemes contain about three times as many dyneins in total than *Chlamydomonas* axonemes. Taken together, the quantitative agreement of ATPase rate ratios with dynein stoichiometric ratio reaffirms the power and utility of our single-cell method.

A further conclusion drawn from our data is that dyneins are the dominant contributor to ATPase activity even for quiescent (inactive) axonemes. A straw-man scenario in which dynein ATPase activity is completely coupled to beating and switched off when the axoneme is quiescent would predict inactive WT and paralyzed WS4 to have nearly the same ATPase activity. On the contrary, our data clearly show that inactive WT has ~ 4X higher ATPase activity than the dynein-deficient WS4. Interestingly, the relatively large spread in ATPase values observed for inactive sperm axoneme data is also reflected in the inactive WT *Chlamydomonas* data, supporting our hypothesis that the spread in inactive ATPase activity stems from varying degrees of mechanical damage.

## Discussion:

### Improvements over previous bulk assays

Our single-cell droplet encapsulation method improves on previous bulk studies in several ways. Most importantly, our measurements are carried out using optical fluorescence in nL volume droplets containing a single cell. This eliminates bulk averaging altogether and permits a direct correlation between the axoneme's beating kinematics and its ATP consumption. A limitation of previous bulk studies is that interpretation of mean ATP consumption rates rely on the assumption of nearly 100% reactivation, whereas even under the most ideal experimental conditions, demembranation and reactivation results in a mixture of nonmotile , motile, and fragmented axonemes (13). This is avoided in our single-cell method which unambiguously measures and directly correlates ATP consumption and beating pattern.  Moreover, in the case of cells with impaired motility, our method permits the consequences of these impairments in motility to be assessed. The method also permits measurements under conditions that were difficult or impossible in previous bulk studies, such as measurements of ATP consumption in high viscosity solutions (Fig. 3F), from axoneme fragments, and from axonemes that transition from beating to non-beating states (Fig. 4). Thus our single-cell method is highly versatile and applicable to a wider range of experimental conditions than previous bulk methods.

In addition to smaller volumes, we have optimized the biochemical conditions for demembranation and reactivation, including the use of an ATP regeneration system to maintain a stable ATP concentration over the entire duration of the measurements. It is important to maintain the ATP concentration because ATP itself has been shown to be an important modulator of dynein activity and ciliary motility (12, 15, 26).

### Causes for the variation in ATP consumption rates of inactive flagella

Our results suggest that there is a wide range of ATP consumption rates for inactive flagella. Cells which have abruptly ceased beating tend to have nearly the same ATP consumption rates as before they stopped (ratio of ATP rate after stopping to motile rate is ~0.85) (Fig. 4), whereas cells which have never been observed to beat seem to have rates that are much lower (ratio of inactive to active ATP rate ~ 0.3) (Fig. S4). Here we speculate on the failure mechanisms that underlie the different ATP consumption rates observed for these two groups of inactive flagella.

One possibility for the difference between these two populations is that the completely inactive cells could have experienced extensive mechanical damage (e.g. extraction of dyneins) during the demembranation and emulsification process, and thus variation in their residual ATP consumption rate reflects differing intrinsic levels of endogenous movement-independent ATPase activity rather than movement-dependent dynein activity. Previous bulk studies utilizing systematically homogenized *S. purpuratus* axonemes also obtained a ~ 0.3 ratio for ATP consumption rate of inactive to active axonemes (13), supporting the damage scenario. By contrast, for the freshly inactive cells in the transitional population, macroscopic coherent beating could have ceased due to a critical structural failure, such as breakage of a doublet microtubule or the links between doublets. Perhaps this smaller degree of damage would still leave dyneins capable of undergoing ATP-consuming powerstrokes and local doublet sliding (without macroscopic beating) and thus the ATP consumption would still occur at nearly the same rate. In this scenario, the variation in non-beating rates reflects the variation in residual movement-dependent dynein activity. Another possible failure mechanism for freshly inactive cells is failure of dynein regulation. This would leave the axoneme in a rigor state with dyneins on both sides active. In this scenario, the axoneme could also be hydrolyzing ATP at nearly the same rate as actively beating axonemes. Overall, the new insight that inactive ATP consumption rates have a wide variation further highlights the critical importance of single-cell measurements.

*The measured ATP consumption rates at single-cell level are consistent with current mechanistic models for flagellar beating*

Our method enabled unambiguous measurement of single-cell flagellar ATP consumption of $(2.3 \pm 0.2) \times 10^5$ ATP/beat. Relating this result to known details about axonemal structure and dynein mechanochemistry allows us to deduce further parameters that shed light on the mechanism of flagellar motility. Kinetic (33-35) and structural (36, 37) models of dynein mechanochemistry consensually assume that one ATP is consumed per powerstroke cycle. Assuming a tight coupling between dynein stepping and doublet microtubule sliding (i.e. no sliding occurs without ATP-actuated dynein attachment/detachment), the number of ATP molecules consumed per beat can be estimated by multiplying the number of active dyneins in the axoneme and the number of steps taken by each active dynein during one complete flagellar beat cycle. Cryo-electron tomography studies have shown that there are ~67,500 dyneins in a sea urchin sperm flagellum (4, 38). A reasonable hypothesis for ciliary and flagellar beating is that not all the dyneins are equally active. Instead, to generate a planar waveform dynein activity should switch periodically between two groups of 2(-3) doublets that lie nearest to the plane containing the central pair. Adopting this assumption, we estimate that there are $67,500 \times 2/9 \sim 15,000$ active dyneins during each beat cycle. We next consider the doublet microtubule displacement that each dynein generates per step; single molecule studies of axonemal dynein suggests an ~8 nm displacement resulting from the dynein power stroke (39, 40). The total sliding displacement along the longitudinal axis of the axoneme between two neighboring doublets is given by the product of axoneme diameter and the shear angle, defined as the interior angle between the symmetry axis of the head and the line tangent to the axoneme immediately after the first bend (41-43). From Fig. 3D, we estimate the shear angle to be ~1 radian. We estimate the effective diameter of the axoneme to be ~ 150 nm (43), giving an interdoublet sliding displacement of ~150 nm. Thus, each dynein takes approximately 150 nm/ 8 nm ~ 19 steps per beat cycle. Assuming each dynein takes 1 step per ATP hydrolyzed, the number of ATPs that are hydrolyzed for 1 complete beat cycle is: 15,000 dyneins × 19 steps/beat ~$2.85 \times 10^5$ ATP molecules, which is in reasonable agreement with the measured value.

A simple energetic estimate further confirms our measured value of consumed ATP/beat, from an entirely different starting point. We consider the axoneme as a classical rigid beam and estimate the amount of energy required to bend it into the roughly sinusoidal shape adopted throughout its active beating cycle. Each beat consumes $2.3 \times 10^5$ ATP which corresponds to $1.84 \times 10^{-14}$ J of energy assuming $20 k_B T$ of free energy for each ATP hydrolyzed to ADP+Pi (44). Previous experiments measured an axonemal bending stiffness of $16.6 \times 10^{-21}$ N m$^2$ for demembranated *Lytechinus Pictus* axonemes in the presence of 0 µM ATP (45). In the presence of 10 mM ATP, similar experiments have shown that axonemal bending stiffness of sea urchin sperm decreases by an order of magnitude (46). Based on the findings of these experiments we estimate a bending stiffness value of $4.5\times10^{-21}$ N m$^2$ for our *L. Pictus* axoneme at the median [ATP] = 20 µM of our experiments. Elastically deforming a rigid beam with this bending stiffness value to curvatures that correspond to the experimentally observed beating patterns requires an energetic cost of ~$9.75\times10^{-15}$ J. This simple estimate does not consider the contribution of hydrodynamic dissipation to the axonemal energy consumption, which would increase the energetic cost and bring the estimate even closer to our measured value for consumed ATP. Thus our measured results are consistent with expectations based on deforming an elastic beam.

For higher viscosity solution we have measured $3.2\times10^5$ ATP/beat, which is ~1.5x higher than the ATP/beat measured in low-viscosity buffer. Our observations suggest that this is a consequence of slower beating rather than increased rate of ATP consumption, since the ATP consumption rate (ATP/sec) is the same for the two viscosities while the beating frequency is lower for higher viscosity fluid. This suggests that the axoneme operates in an "open-loop" mode, i.e. it draws a fixed input power despite increased external load. The 1.5x ratio of the ATP/beat is consistent with the shear angle ratio (~1.57 radian / 1 radian) of the high to low viscosity axonemes in Figs. 3E and 3D respectively. Since the ratio of shear angle is exactly equal to the ratio of sliding displacements, this suggests that each dynein is taking ~1.5x more steps during a beat cycle in high viscosity solution. Experiments with reactivated sea urchin flagella in water in which nexin links were partially digested with elastase resulted in higher curvature beating waveforms that are very similar to what we observe in high viscosity solution (47). This supports the picture that increased waveform curvature results from increased doublet sliding displacement induced by dyneins taking more steps during each beat cycle. The increased step number (and hence sliding displacement) for higher curvature waveforms is also consistent with models in which axonemal bending is generated from constrained inter-doublet sliding (11, 27, 41). Taken together, our viscosity-dependent measurements reveal that increasing the viscosity of the background fluid directly impacts the regulation of dynein activity in the axoneme.

***The energy input to the axoneme is primarily expended on elastic deformation rather than overcoming viscous dissipation***

There are two channels for energy expenditure in the axoneme: overcoming viscous drag and generating elastic deformation of the axoneme. Sources of viscous dissipation include hydrodynamic drag and molecular friction; sources of elasticity include bending of the axoneme as well as stretching of elastic links (such as the nexin linker) that resist sliding and convert sliding into bending (1). To interpret the linear dependence of the ATP consumption rate on the beating frequency of the axoneme (Fig. 3C, H), we need to understand the energetic contributions from viscous dissipation and elasticity and how these contributions scale with frequency. Modeling the beating flagellum as an Eulerian beam with active bending moments immersed in a viscous fluid yields an expression for the power expended in generating sinusoidal undulations (48):

$$P = b^2 L \left[ \frac{\beta}{2}(2\pi f)^2 + \alpha f \left(\frac{2\pi}{\lambda}\right)^4 \right] \qquad (1)$$

where L is the length of the flagellum and $b$, $\lambda$, $f$ are the amplitude, wavelength, and frequency of a propagating sine wave $y(x,t) = b*\sin[2\pi(x/\lambda - ft)]$ approximating the beating waveform. The resistive force theory approximation for the transverse hydrodynamic drag per unit length for motion perpendicular to the flagellum axis is given by $\beta = 4\pi\eta/[\log(2\lambda/a) + 0.5]$, where $\eta$ is the viscosity of the fluid, $a$ is axoneme radius, and $\alpha = EI$ is the bending stiffness of the flagellum ($E$ is the Young's modulus, $I$ is the area moment of inertia).

The frequency dependence of our ATP consumption data in Figure 3C, H can be readily interpreted with Eq. (1). The first term, scaling as $f^2$, represents the contribution of the hydrodynamic drag to the power. The second term, scaling linearly with $f$, represents the contribution of bending elasticity to the power. Our ATP consumption data clearly scales linearly with frequency for both low and high viscosity (Fig. 3C, H), suggesting that the ATP energy input into the axoneme is channeled primarily into elastic deformation, rather than overcoming hydrodynamic drag.

To understand why most of the energy is channeled into elastic deformation, it is useful to estimate the relative magnitude of the terms in Eq. (1). Estimating the relevant parameters from the nearly sinusoidal high-viscosity waveform (Fig. 3E) we assume that $b = 3.5$ μm, $\lambda = 21$ μm, $L = 45$ μm, $f = 4$ Hz, $\beta = 0.035$ Pa s, and $\alpha = 4.5 \times 10^{-21}$ N m$^2$ (45, 46). Using this information we find that energy expenditure associated with the elastic deformation is $\sim 8.1 \times 10^{-14}$ J/s, which is approximately an order of magnitude larger than the hydrodynamic dissipation term $\sim 6.4 \times 10^{-15}$ J/s. A similar estimation for the low viscosity waveform also reaches the conclusion that the elastic term is at least an order of magnitude larger than the hydrodynamic term. Thus, elastic deformation of the axoneme requires an order of magnitude more energy than overcoming hydrodynamic drag.

By comparison, our data in Figure 3H yield an estimate of the power input from ATP hydrolysis $\sim 1.1 \times 10^{-13}$ J/s for f = 4 Hz, based on the assumption that hydrolysis of one ATP yields 20 $k_B T$ of free energy. In principle there could be other energetic terms arising from inter-doublet sliding elasticity and friction (49), but these terms would have similar frequency dependence to bending and hydrodynamic dissipation, respectively. For the case of inter-doublet sliding elasticity, experimental measurements of the internal sliding resistance in *Chlamydomonas,* sea urchin, and rat axonemes yield a consensus value of $\sim 0.02 - 0.1$ mN/m per 96 nm repeat (50-52). Assuming a typical sliding displacement of $\sim 150$ nm (43), this yields a sliding power expenditure $\sim 1 \times 10^{-15}$ J/s at f = 4 Hz, which is 2 orders of magnitude smaller than the bending elasticity contribution. Thus we are justified in neglecting elastic terms arising from internal sliding elasticity in Eq. (1).

In theory, the hydrodynamic term should dominate at higher viscosity and frequency, but we were unable to obtain measurements at viscosity higher than 8x water due to the prohibitively low success rate of reactivated axonemal motility at higher viscosities. The fraction of motile sperm dropped sharply for concentrations above 0.5% methyl cellulose (13). A compounding effect may be damage to the axoneme caused by high shear stresses during the emulsification process. Future studies may employ the use of temperature-sensitive yield stress fluids to circumvent shear damage during emulsification in order to achieve encapsulation experiments at higher viscosities to resolve the crossover from elastic to viscous dominated regime.

*Overall swimming efficiency of axonemes increases with viscosity*

The overall swimming efficiency of an organism at low Reynolds number can be characterized by a ratio of how much power it expended to move a certain distance to how much power it consumed to move. It follows that this overall efficiency, $\varepsilon_{swim}$, is the product of the swimmer's chemomechanical efficiency $\varepsilon_{chemo}$ (the ratio of mechanical power $P_{mech}$ the swimmer expended to generate its mechanical stroke pattern to its rate of chemical (ATP) energy consumption $P_{ATP}$) and its hydrodynamic efficiency $\varepsilon_{hydro}$ (the ratio of the power $P_{drag}$ required to translate the swimmer with a constant shape through the fluid at its average velocity to the mechanical power the swimmer expended to generate its mechanical stroke pattern $P_{mech}$) (53, 54). Recent studies of biological swimmers have focused almost exclusively on $\varepsilon_{hydro}$ as a measure of efficiency, rather than $\varepsilon_{swim}$ (49, 55, 56). This is partly due to the lack of robust experimental methods for determining $\varepsilon_{chemo}$, which requires determination of both ATP consumption at a molecular level and the beating kinematics of the axoneme.

In comparison, our method enables determination of the axoneme $\varepsilon_{chemo}$. Supplemental table S1 summarizes all relevant parameters that were extracted from the beating kinematics to calculate the efficiencies in low and high viscosity solutions for sperm reactivated with [ATP] = 20 μM. The mechanical power expended by the axoneme to generate its beating waveform, $P_{mech}$, is computed from experimentally measured quantities using Eq. (1), whereas the rate of chemical energy consumption, $P_{ATP}$, is determined from the measured value of ATP consumption rate for [ATP] = 20 μM (Fig. 3A,F). Dividing these two numbers yields $\varepsilon_{chemo}$ = 0.34 for a low-viscosity suspension and $\varepsilon_{chemo}$ = 0.6 for a high-viscosity suspension. We also determine the hydrodynamic efficiency $\varepsilon_{hydro}$ as the ratio of $P_{drag}$ to $P_{mech}$ with $P_{drag} = (\beta/2)LU^2$ where U is the mean velocity of the sperm, determined by tracking the swimming sperm's displacement as a function of time for a sperm reactivated at [ATP] = 20 μM in low and high viscosity, where $\beta/2$ is the tangential component of the drag coefficient from resistive force theory for a flagellum of length L (48). Surprisingly, we find that both $\varepsilon_{chemo}$ and $\varepsilon_{hydro}$ of the flagellum beating in high viscosity buffer are higher than in low-viscosity, with a larger difference in hydrodynamic efficiency. Consequently, the overall axoneme swimming efficiency is about an order of magnitude higher for sperm swimming in high viscosity ($\varepsilon_{swim}$ = .008) than low viscosity ($\varepsilon_{swim}$ = .001), a somewhat counterintuitive result given that sea urchin sperm have presumably evolved for optimum motility in low-viscosity sea water. Future studies will be necessary to map out the subtle interplay of chemomechanical and hydrodynamic efficiencies over a wider range of fluid viscosity and viscoelasticity.

## Conclusion:

We have described a new experimental platform that enables simultaneous visualization of flagellar beating patterns and single-cell measurements of the number of ATP molecules consumed per flagellar beat cycle. These single-cell measurements have contributed significant insights into the mechanisms of active force generation in the axoneme. Most important is the predominant contribution of elastic deformation over hydrodynamic dissipation in energy consumption. In addition, we have obtained a measurement of chemomechanical efficiency and its dependence on fluid viscosity. In particular, we have shown that overall swimming efficiency increases with viscosity, with a counterintuitive dependence on fluid viscosity. Coupled with theoretical modeling, our high resolution measurements will be useful in discriminating between putative mechanisms of beating regulation in cilia and flagella. With slight modifications, our methods could be fruitfully applied to gain molecular level insight into several interesting problems in flagellar mechanics, including energy dependence of flagellar synchronization and beating efficiency in physiologically relevant viscoelastic fluids. Finally, the method could also be used for energetic analysis of other far-from-equilibrium assemblies such as synthetic cilia, active emulsions, and vesicles (57-59).

--------------

**Author Contributions:** DC, DN, SF, and ZD designed research; DC and MH performed research; MH and SF contributed experimental tools; DC analyzed data; DC, DN, and ZD wrote the manuscript.

**Acknowledgments:** We are grateful to Tom Powers for stimulating discussions. We thank Kangkang Song for suggesting and obtaining *Chlamydomonas* WS4 mutant strain, Win Sale for providing the WS4 strain, and Gang Fu for preparation of *Chlamydomonas* WT and WS4 axonemes. This research was supported by the National Science Foundation (MCB-1329623 to ZD), MRSEC (NSF-MRSEC-1420382, NSF-MRSEC-1206146), the National Institutes of Health (GM083122 to DN) and the W.M. Keck foundation (to ZD, DN, and SF). We also acknowledge use of MRSEC optical microscopy and microfluidic facilities. The authors declare no conflicts of interest.

Figures:

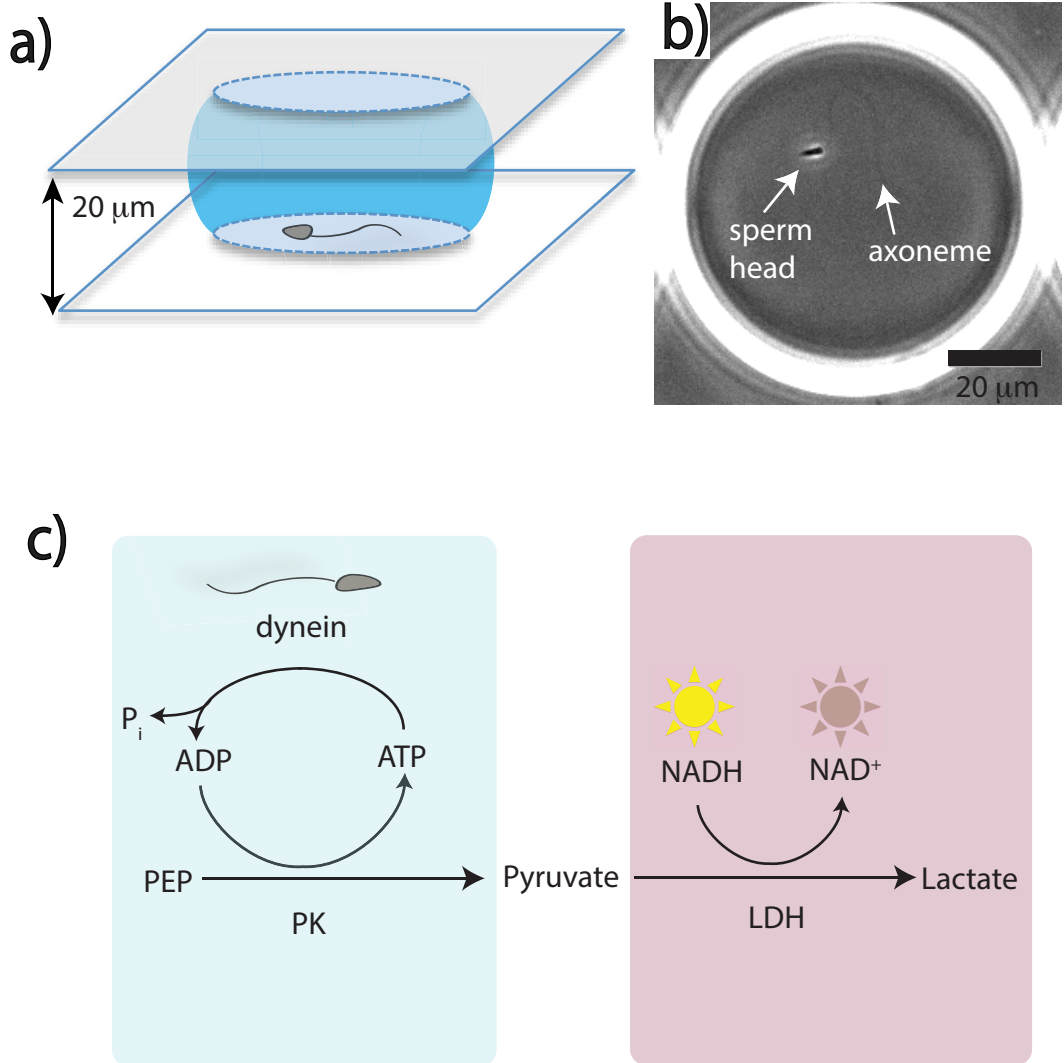

**Figure 1**: Schematic of emulsion droplet geometry and biochemistry. (a) Compressed emulsion droplet schematic. Droplet contains aqueous phase (blue) surrounded by oil phase (uncolored). For details refer to Materials and Methods section in the main text. (b) Phase contrast image of sperm with undulating axoneme in droplet (see also Supplementary Movie S1). (c) Biochemistry of coupled ATP regeneration and NADH fluorescence reporter system. See Materials and Methods section of the main text for details on the biochemical components.

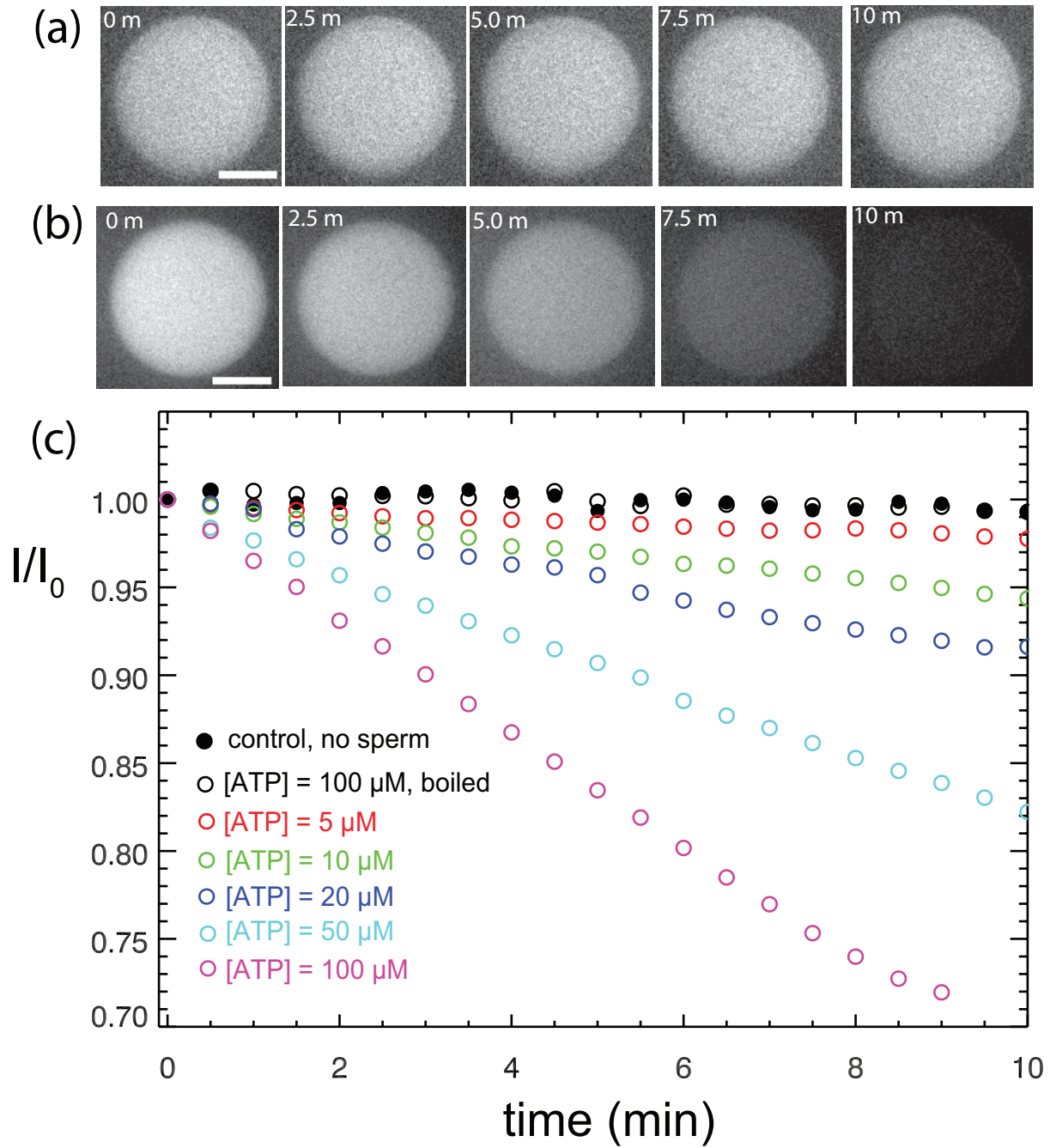

**Figure2**: NADH fluorescence from droplets. (a) Series of snapshots of droplet fluorescence over a 10 minute period from a droplet containing boiled and thus enzymatically inactive sperm. No obvious intensity drop is observed over time. Scale bar is 40 µm. (b) Series of snapshots of droplet fluorescence over a 10 minute period from a droplet containing a single sperm with actively beating flagellum; [ATP] = 100 µM. (c) Representative data for normalized intensity of droplet NADH fluorescence as a function of time for various [ATP] and controls (no sperm and boiled sperm, respectively)

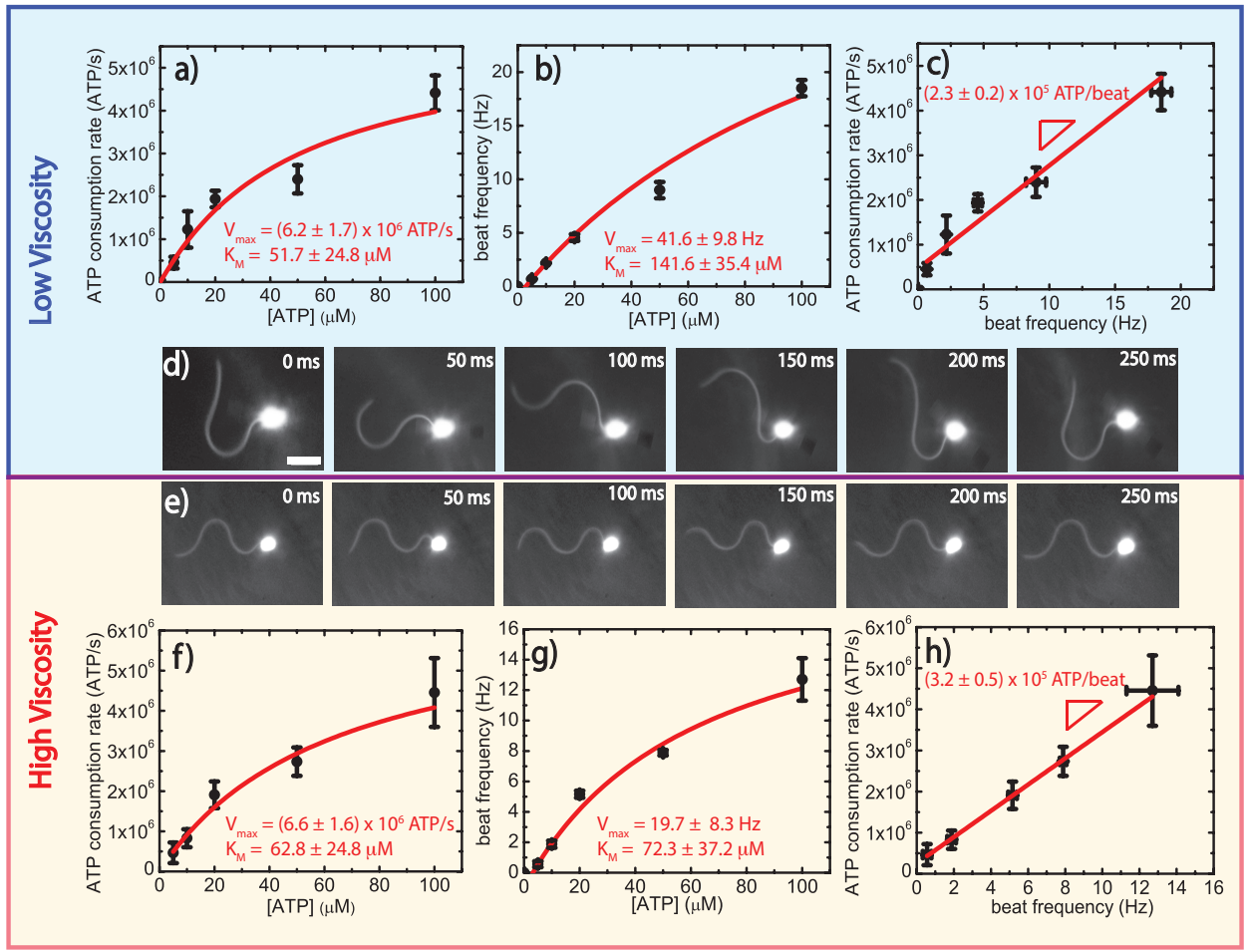

**Figure 3**: Data for ATP consumption rate and beat frequency measurements for sperm in reactivation buffer (low viscosity) (a-d) and in reactivation buffer with 0.5% w/v methyl cellulose (high viscosity) (e-h), respectively. (a) Mean ATP consumption rate as a function of [ATP]. Error bars are ± standard error of the mean (SE) (N=4). Line is a fit to Michaelis-Menten relation: $V_{max}[ATP]/(K_M + [ATP])$. (b) Mean beat frequency as a function of [ATP] with modified Michaelis-Menten fit (see text). Error bars are ± SE (N=4). (c) ATP consumption rate plotted as a function of beat frequency (same data as a-b). Line is least-squares fit to the data, with slope yielding ATP/beat. (d) Darkfield images of beating waveform over one complete beat cycle for demembranated *Lytechinus pictus* sperm that was reactivated in buffer with 20 μM ATP. Scale bar is 10 μm. (e) Darkfield images of beating waveform over one complete beat cycle for demembranated *Lytechinus pictus* sperm that was reactivated in high viscosity buffer containing 0.5% w/v Methyl Cellulose and 20 μM ATP. The time intervals and spatial scales are identical to panel in d. (f) Mean ATP consumption rate as a function of [ATP] with modified Michaelis-Menten fit (see text). Error bars are ± SE (N=4). (g) Mean beat frequency as a function of [ATP] with Michaelis-Menten fit. Error bars are ± SE (N=4). (h) ATP consumption rate plotted as a function of beat frequency (same high-viscosity data as f-g). Line is least-squares fit, with slope yielding ATP/beat.

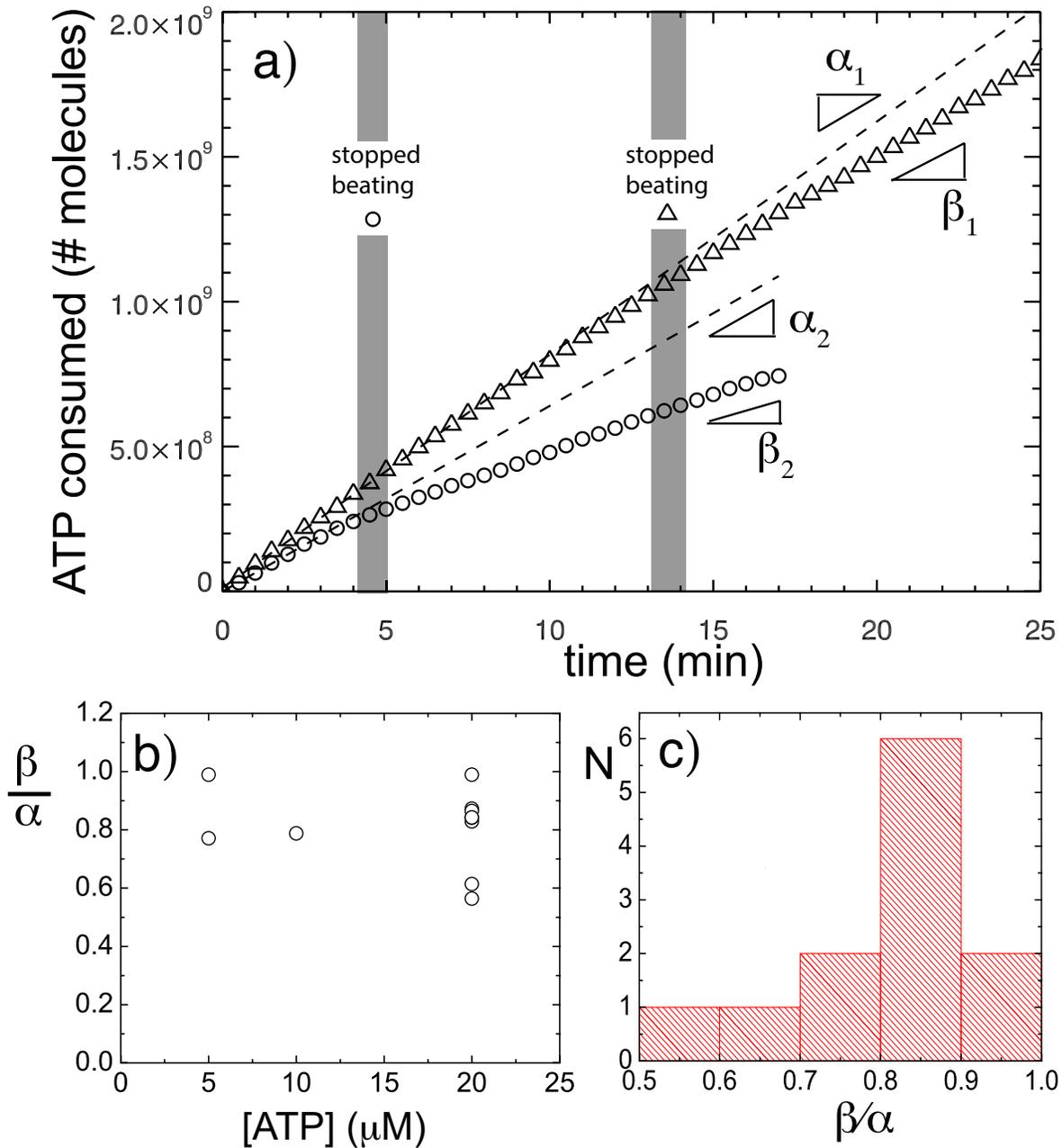

**Figure 4**: Beating to non-beating transition captured in single-cell droplet time series. (a) Plot of ATP consumed vs. time elapsed during measurement for two different sperm (circles and triangles), both in the presence of 20 μM ATP. The time interval during which beating ceases is indicated in grey. Dashed line is linear fit to the data over the actively beating portion of the time series. $\alpha$, $\beta$ are the ATP consumption rates before and after beating stops, respectively, determined from linear fits of the data. For the two different sperm: $\beta_1/\alpha_1 = 0.84$, $\beta_2/\alpha_2 = 0.61$, respectively. (b) Ratio of non-beating to beating ATP consumption rate $\beta/\alpha$ as a function of [ATP] for all data exhibiting a beating to non-beating transition (N = 12). (c) Histogram of $\beta/\alpha$ values (N = 12) for the beating to non-beating transition data. The measured values range between 0.5 - 1.0 with a pronounced peak at ~0.85.

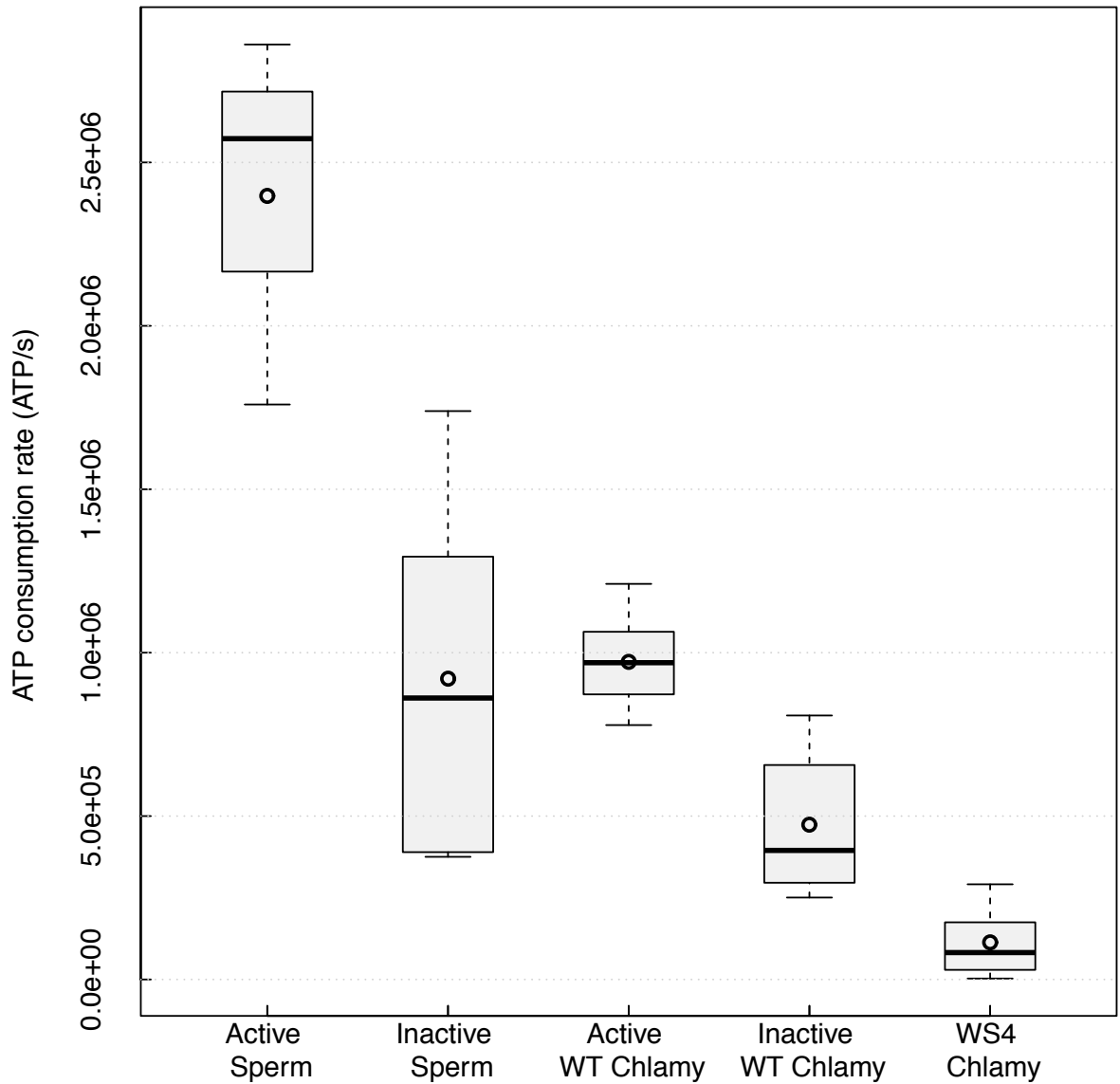

**Figure 5**: Boxplot of ATP consumption rate measured for active sperm (N = 3), inactive sperm (N=6) , active WT *Chlamydomonas* axoneme (N= 10), inactive WT *Chlamydomonas* axoneme (N=10) and dynein-deficient *Chlamydomonas* WS4 mutant (N= 10). All experiments were performed in reactivation buffer with [ATP] = 50 μM (see Materials and Methods section). Thick line denotes the sample median of the distribution and circle is the sample mean. Lower and upper boundaries of the shaded box region denote the first and third quartile, respectively. Whisker boundaries denote minimum and maximum values for each distribution.

**Supplementary Materials**

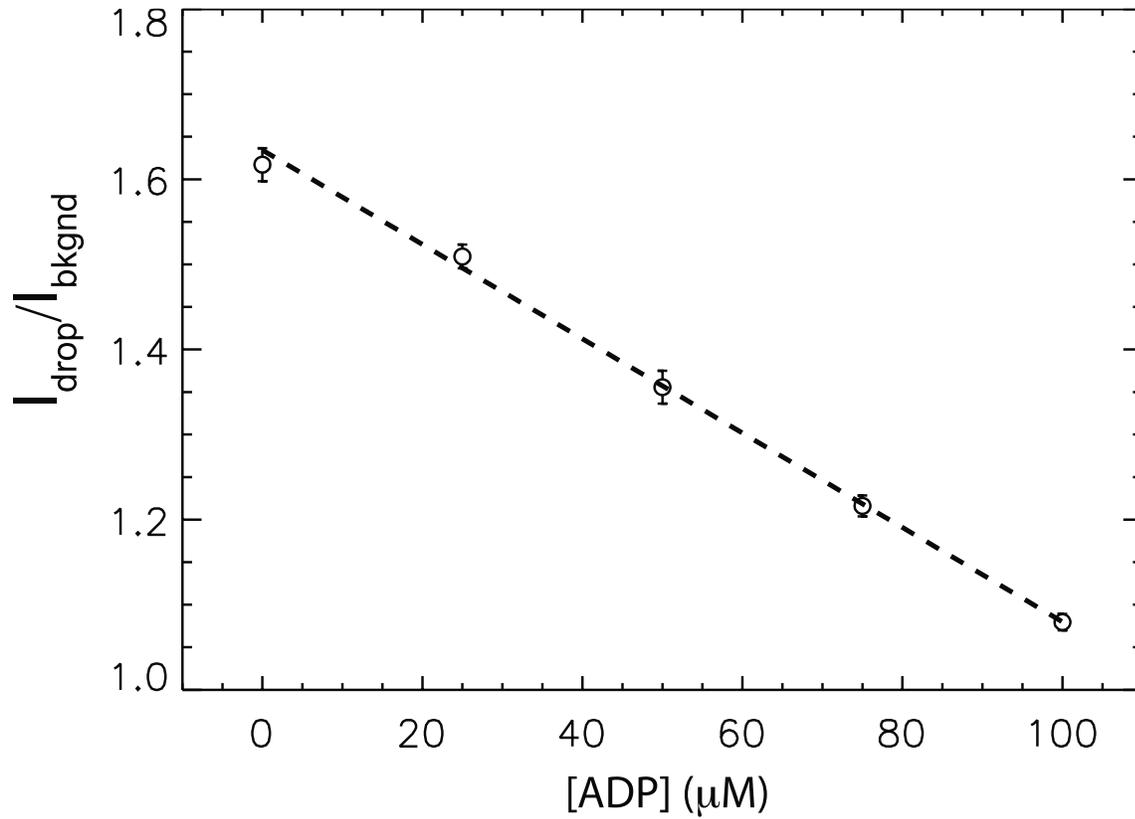

**Figure S1**: Calibration Curve of NADH fluorescence in droplet without sperm cell. Plot of mean intensity ratio of drop to background vs. [ADP]. Error bars are ± SEM (n=4). The dashed line is a least squares fit to the data.

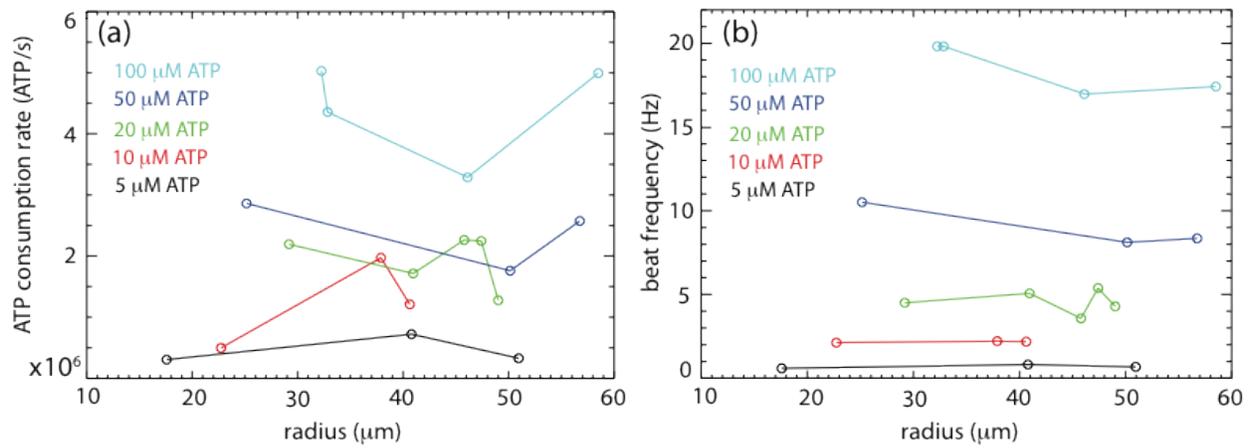

**Figure S2**: The analyses of droplet radius dependence in relation to ATP consumption rate (a) and beating frequency (b) for various [ATP] show absence of systematic trends with droplet size. Lines are eye-guides.

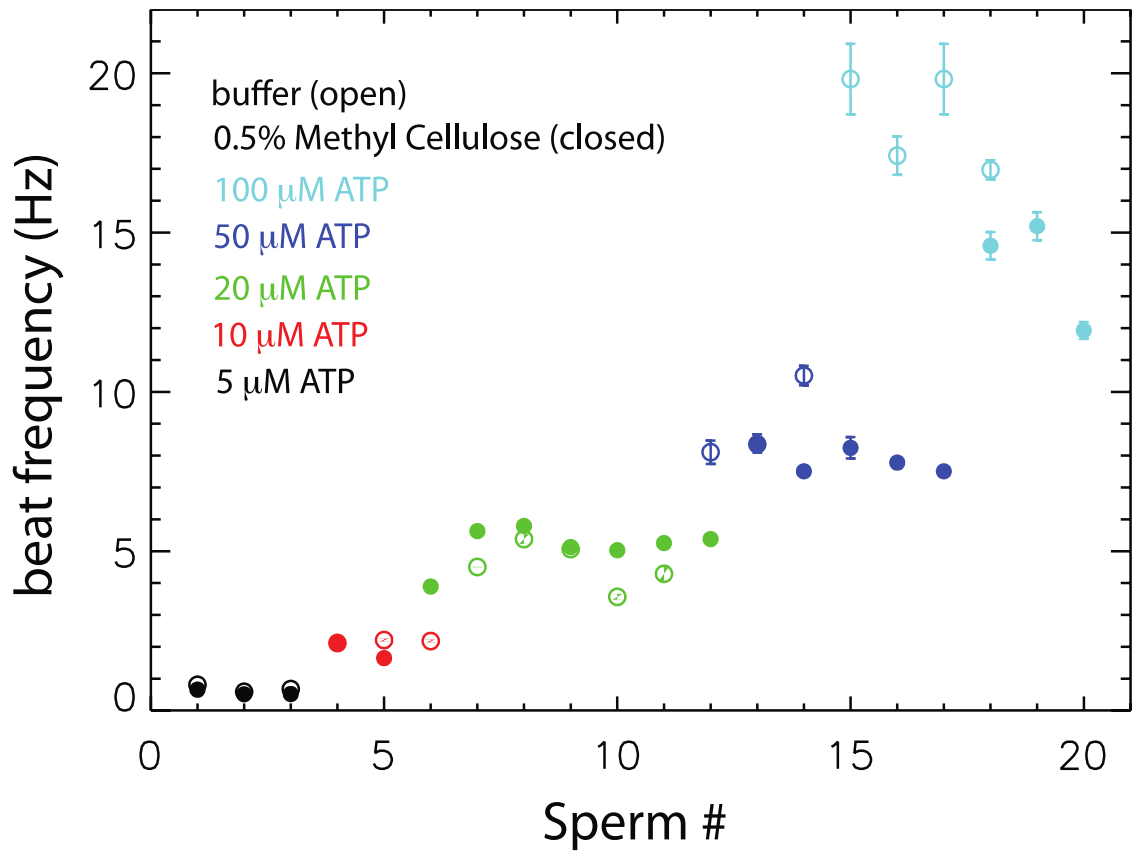

**Figure S3**: Measured beat frequency for each sperm in the dataset. Each point is the mean beat frequency ± SEM for n=5 independent measurements of the beating frequency at different times. Open circles are for sperm in standard buffer; filled circles are sperm in 0.5% w/v methyl cellulose (high viscosity). Error bars are smaller than the scatter between individual sperm at each concentration showing that inter-sperm variation is large relative to experimental resolution of the beating frequency.

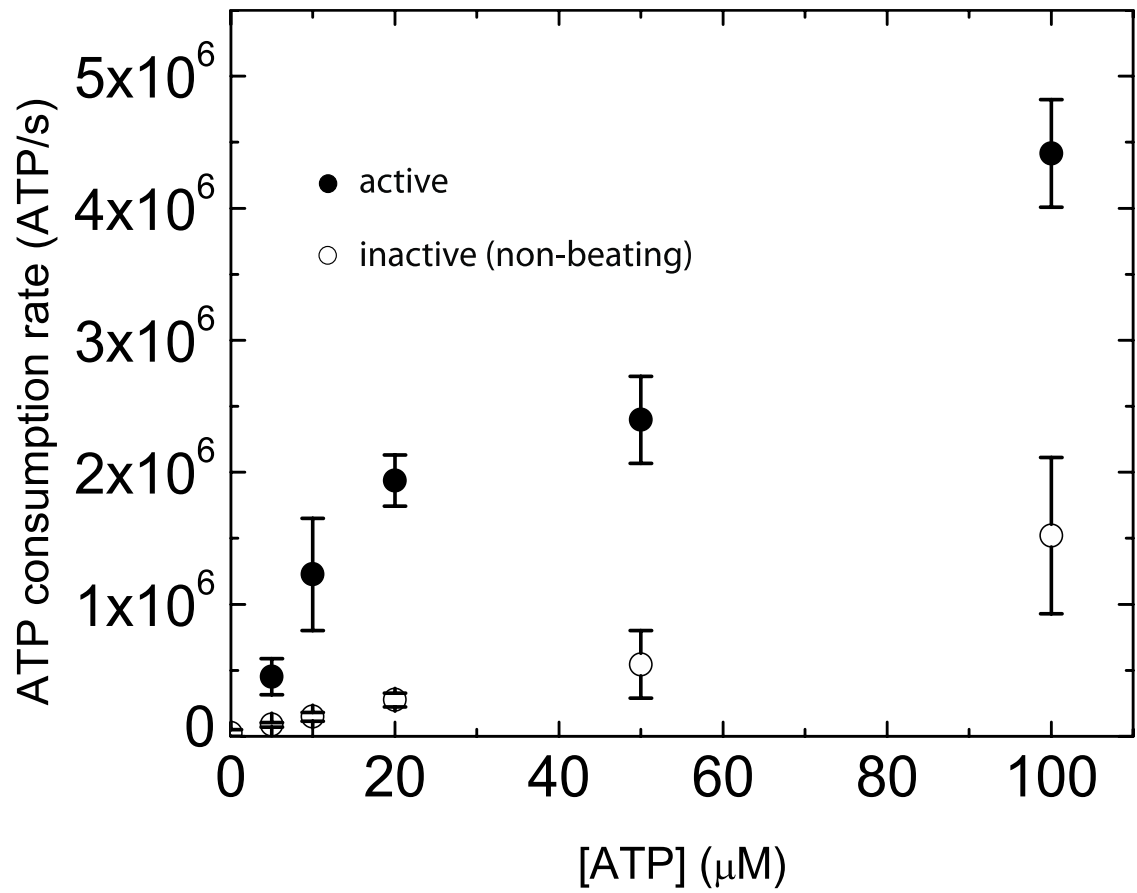

**Figure S4**: Mean ATP consumption rate vs. [ATP] for active and inactive (non-beating) sperm. Error bars are ± SEM (n=4).

| | b (μm) | L (μm) | β (mPa s) | α (N m²) | λ (μm) | f (Hz) | U (μm s⁻¹) | $P_{mech}$ (J s⁻¹) | $P_{drag}$ (J s⁻¹) | $P_{ATP}$ (J s⁻¹) | $\varepsilon_{chemo}$ | $\varepsilon_{hydro}$ | $\varepsilon_{swim}$ |
|---|---|---|---|---|---|---|---|---|---|---|---|---|---|
| Low Viscosity | 5 | 45 | 4 | 4.5 ×10⁻²¹ | 30 | 5 | 46.4 | 5.25 ×10⁻¹⁴ | 1.94 ×10⁻¹⁶ | 1.52 ×10⁻¹³ | 0.34 | .004 | .001 |
| High Viscosity | 3.5 | 45 | 35 | 4.5 ×10⁻²¹ | 21 | 4 | 38.7 | 9.15 ×10⁻¹⁴ | 1.18 ×10⁻¹⁵ | 1.52 ×10⁻¹³ | 0.6 | .013 | .008 |

**Table S1**: Values of beating waveform parameters in Eq. (1) used to calculate various measures of efficiency for low and high viscosity beating waveforms for [ATP] = 20 μM sperm in buffer (low-viscosity) and 0.5% Methyl Cellulose (high-viscosity).

**Movie S1**: Movie of sperm swimming in droplet, [ATP] = 5 μM. Scale bar is 20 μm.

**Movie S2**: Movie of NADH fluorescence in two neighboring droplets. The left droplet does not contain a sperm cell, whereas the right droplet contains one actively swimming sperm. [ATP] = 50 μM. Scale bar is 20 μm.